\documentclass[twocolumn,prl,preprintnumbers,tightenlines,nofootinbib,superscriptaddress]{revtex4-2}

\pdfoutput=1
\usepackage{amsmath,amssymb,amsfonts}
\usepackage{graphicx}
\usepackage[normalem]{ulem}
\usepackage{slashed}
\usepackage{booktabs}
\usepackage{hyperref}
\usepackage{color}
\usepackage[sort&compress]{natbib}
\usepackage{tikz}
\usepackage{cleveref}
\usepackage{cancel}
\newcounter{qnumber}

\newcommand{\chpt}{$\chi$PT}

\newcommand{\nc}{\newcommand}
\nc{\beq}{\begin{equation}}  \nc{\eeq}{\end{equation}}
\nc{\bealg}{\begin{equation}\begin{aligned}} 
\nc{\eealg}{\end{aligned}\end{equation}} 

\newcommand{\tr}{\mathrm{tr}}
\newcommand{\pd}{\partial}
\newcommand{\Uc}{U^\dagger}
\newcommand{\chic}{\chi^\dagger}

\newcommand{\Eq}[1]{Eq.~(\ref{#1})}

\begin{document}

\title{Dynamical Up-quark Mass Generation in QCD-like theories}

\author{Csaba Cs\'aki}
\email{csaki@cornell.edu}
\affiliation{Laboratory for Elementary Particle Physics, Cornell University, Ithaca, NY 14853, USA}

\author{Tuhin S. Roy}
\email{tuhin@theory.tifr.res.in}
\affiliation{Department of Theoretical Physics, Tata Institute of  Fundamental Research, \\ Homi Bhabha Rd, Mumbai 400005, India}

\author{Maximilian Ruhdorfer}
\email{m.ruhdorfer@stanford.edu}
\affiliation{Stanford Institute for Theoretical Physics, Stanford University, Stanford, CA 94305, USA}

\author{Taewook Youn}
\email{taewook.youn@cornell.edu}
\affiliation{Laboratory for Elementary Particle Physics, Cornell University, Ithaca, NY 14853, USA}
\affiliation{School of Physics, Korea Institute for Advanced Study, Seoul 02455, Republic of Korea}

\begin{abstract} 
We calculate the dynamically generated up quark mass in some QCD-like theories with $F=3$ light flavors, obtained from supersymmetric QCD perturbed via anomaly mediated supersymmetry breaking. We match the low-energy effective theory to the traditional chiral Lagrangian of QCD and determine the coefficients to next-to-leading order in chiral perturbation theory, while also varying the number of colors $N$. We find that the dynamically generated up quark mass vanishes in the large $N$ limit, and is small for $F<N$, however for $F=N$ there is a sizeable ${\cal O}(1)$ contribution. While our results are reliable only for small supersymmetry breaking, we observe that extrapolating the $F=N$ result to large supersymmetry breaking  would lead to a  dynamical up quark mass that is large enough to account for its entire physical mass.

\end{abstract}

\maketitle

\section{Introduction}
\begin{figure}[h]
  \centering
  \includegraphics[width=0.7\columnwidth]{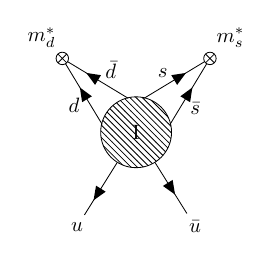}
  \caption{Feynman diagram for the 't~Hooft operator corresponding to the instanton induced up quark mass.}
  \label{fig:inst}
\end{figure}

The strong CP problem of QCD is one of the most puzzling open questions in the standard model (SM) of particle physics. The $\theta$-parameter of QCD explicitly violates CP and feeds into the expression for the electric dipole moment (EDM) of the neutron, which can only be reconciled with experimental bounds if $\bar\theta < 10^{-10}$, even though a priori any value of $\bar\theta$ seems equally possible. The simplest resolution to this puzzle would arise if one of the quark masses (in practice the up-quark mass) were to vanish. This would provide an anomalous, but classically unbroken, $U(1)$ axial symmetry that could be used to rotate away the $\bar \theta$ angle and render it unphysical. While this solutions seems theoretically very appealing, the physical up quark mass appears to be non-vanishing. However, it was pointed out already by Kaplan and Manohar in 1986~\cite{Kaplan:1986ru} that dynamical effects contributing to the pion mass can effectively shift the up quark mass, possibly allowing a successful fit to the pseudoscalar meson masses even if the underlying up quark Yukawa coupling was vanishing. One simple interpretation of this is that non-perturbative effects generate a dynamical quark mass contribution, which corresponds to an additive renormalization of the quark mass. One often cited example would be instanton contributions to the up-quark mass in a three flavor theory, as depicted in Fig.~\ref{fig:inst}. Such an instanton would provide corrections $\Delta m_u \propto m_d^* m_s^*$, and could be sizeable. 
Of course the instantons are strongly coupled and uncalculable using standard instanton calculus (see e.g.~\cite{Csaki:2023ziz} for simple power counting rules to estimate instanton effects). The important point is that the strong CP problem would still be solved if all of the up quark mass was dynamically generated, as long as the underlying up Yukawa coupling was vanishing. 

Generally it is very difficult to calculate the dynamical quark masses from first principles due to the strong dynamics. The best route is via lattice simulations (see e.g.~\cite{Fodor:2016bgu,FlavourLatticeAveragingGroupFLAG:2024oxs} for an overview), though there is some ambiguity to disentangle the underlying up quark mass from the dynamical corrections~\cite{Frison:2016rnq}. Another possibility is to use phenomenological fits of the low-energy constants in the chiral Lagrangian to experimental data~\cite{Gasser:1984gg,Bijnens:2011tb,Bijnens:2014lea,Dine:2014dga}. However, the current precision is not sufficient to significantly narrow down the dynamical contribution to the up quark mass. Finally large-$N$ QCD allows to obtain estimates for the low-energy constants in the chiral Lagrangian~\cite{Leutwyler:1996sa,Kaiser:2000gs,Davies:2022ueb}. In particular Ref.~\cite{Davies:2022ueb} recently determined the low-energy constants to leading order in the large-$N$ expansion, finding that the dynamically generated contribution to the up quark mass is small in the large-$N$ limit.

In this paper we calculate the dynamical quark mass corrections in a QCD-like theory obtained from a supersymmetric (SUSY) version of QCD with three flavors ($F=3$), where SUSY is broken via anomaly mediated supersymmetry breaking (AMSB). As shown in~\cite{Murayama:2021xfj,Csaki:2022cyg}, this theory has the right massless degrees of freedom in the UV (ie. the squarks and gluinos get masses proprtional to the SUSY breaking scale $m$), while there is a QCD-like vacuum (which in many cases is the global minimum of the potential, depending on the number of flavors and colors) in which the $SU(3)_L\times SU(3)_R$ flavor symmetry is broken to $SU(3)_V$. The corresponding Goldstone bosons are massless in the limit of vanishing quark masses, and a systematic evaluation of the chiral Lagrangian is possible and mostly calculable, up to a few uncalculable constants for the case when $F=N$. 

In this work, we determine the chiral Lagrangian for these QCD-like models to second order in the quark masses (which for $F=3$ should capture the leading effects of dynamical up-quark mass generation) and to fourth order in derivatives. This will entail the matching of a linear sigma model obtained from the SUSY theory to the non-linear theory of $\chi$PT. Using our second order chiral Lagrangian we will determine the dynamically generated up quark mass. We find that the coefficient $L_4$ and $L_6$ (which as we will explain contains the information on the dynamical up quark mass) is unsuppressed for finite $N,F$ compared to the other coefficients in the chiral Lagrangian. If one were to extrapolate to large SUSY breaking $m/\Lambda \sim {\cal O}(1)$ (beyond the validity of our approximations) one would numerically obtain large dynamical up-quark mass corrections for the $F=N=3$ case that could be accounting for the entire observed up quark mass, while for $F<N$ we find smaller corrections. 

\section{Dynamical up quark mass from  $\chi$PT}
First we review how the dynamical quark mass generation is incorporated in traditional $\chi$PT. The leading order \chpt~Lagrangian is simply given by
\beq\label{eq:LOChiralLag}
\mathcal L_0 = \frac14 F_0^2 \, \bigg\{ \tr [\pd^\mu U^\dagger \pd_\mu U] + \tr[\chic U + \chi U^\dagger] \bigg\},
\eeq
where $U$ is the unitary three-by-three matrix containing the eight pseudo-scalar Goldstone bosons, $F_0$ is the leading-order pion decay constant and $\chi = 2 B_0 M_Q$ with $B_0 = -\langle \bar{q} q\rangle /(3 F_0^2)$  related to the quark condensate~\cite{Gell-Mann:1968hlm}. Here $\chi$ does not depend on the renormalization scale and scheme, while $B_0$ and $M_Q$ do. This Lagrangian will determine the leading 4-pion interactions as well as the usual Gell-Mann Okubo (GMO) mass relations. However as we saw in the introduction, there are dynamical contributions to the effective mass which scale like $\Delta m_u\propto m_d^* m_s^*$, implying that one needs to include terms which are higher order in the quark masses or derivatives in the chiral Lagrangian. In the absence of external gauge fields the NLO chiral Lagrangian is comprised of terms with either two powers of the quark mass, one power of the quark mass and two derivatives, or four derivatives. $\chi$PT traditionally uses the following basis of eight operators~\cite{Gasser:1984gg}
\beq \label{eq:NLOChiralLag}
\mathcal L_1 = \sum_{i=1}^{8} L_i \mathcal O_i,
\eeq
where
 \begin{widetext}
\bealg
\mathcal O_1 &= \tr[\pd_\mu U \pd^\mu U^\dagger]^2, &\,
\mathcal O_2 &= \tr[\pd_\mu U \pd_\nu \Uc ] \tr[\pd^\mu U \pd^\nu \Uc] \\
\mathcal O_3 &= \tr[(\pd_\mu U \pd^\mu \Uc)^2 ], &\,
\mathcal O_4 &= \tr[\pd_\mu U \pd^\mu \Uc] \tr[\chic U + \chi \Uc] \\
\mathcal O_5 &= \tr[\pd_\mu U \pd_\mu \Uc(\chic U + U \chi)], &\,
\mathcal O_6 &= \tr[\chic U + \Uc \chi]^2 \\
\mathcal O_7 &= \tr[\chic U - \Uc \chi]^2, &\,
\mathcal O_8 &= \tr[\chic U \chic U + \chi \Uc \chi \Uc] \\
\eealg
\end{widetext}

These ${\mathcal O}_{1,\ldots ,8}$ operators form a complete set and  capture all non-perturbative effects including for example instanton contributions. To see this, consider the effective operator generated by the instanton diagram in Fig.~\ref{fig:inst} 
\begin{equation}
\label{eq:OpInstanton}
\mathcal O_{-1} = \tr[{\rm cof}\, \chi^\dagger U^\dagger ] + h.c.,
\end{equation}
where the cofactor is given by ${\rm cof}\, \chi^\dagger = \det (\chi^\dagger) \ (\chi^\dagger)^{-1}$.  As expected, this operator is not independent of the ${\mathcal O}_{1,\ldots ,8}$ set. The easiest way to see this is by using the Kaplan-Manohar identity for $SU(3)$ matrices~\cite{Kaplan:1986ru}
\begin{equation}
\tr[{\rm cof}\, \chi^\dagger U^\dagger ] = \frac{1}{2} \bigg\{\tr[\chi^\dagger U]^2 - \tr[\chi^\dagger U \chi^\dagger U]\bigg\} ,
\label{eq:KaplanManohar}
\end{equation}
which implies that
\begin{equation}\label{eq:instantonOperator}
\mathcal O_{-1} = \frac{1}{4} (\mathcal O_6 + \mathcal O_7) - \frac{1}{2} \mathcal O_8.
\end{equation}
The upshot is that all of the QCD dynamics (to the given order in expansion in the quark masses and derivatives) is encoded in the coefficients $L_{1,\ldots , 8}$. 

In fact the Kaplan-Manohar identity (\ref{eq:KaplanManohar}) allows one to further simplify the Lagrangian: it can be utilized as a gauge transformation on the $\chi$PT Lagrangian at $\mathcal{O}\left(p^4\right)$. The physics remains invariant under the shift of the mass matrix~\cite{Kaplan:1986ru} as long as some of the coefficients $L_{1,\ldots ,8}$ are also shifted according to be following transformation: 
\begin{equation}
    \begin{split}
        \chi \rightarrow \chi + a\: {\rm cof}(\chi^\dagger ) \, , \quad  
     L_6 \rightarrow L_6 - a \: \frac{F_0^2}{16} \, \\
     L_7\rightarrow L_7 -a \: \frac{F_0^2}{16} \, , \quad \text{and} \quad
     L_8\rightarrow L_8 + a \: \frac{F_0^2}{8}  \, , 
    \end{split}
\end{equation}
for arbitrary $a$.
One way of utilizing  this redundancy is to eliminate the instanton generated opearator in Eq.~\eqref{eq:OpInstanton} by choosing $a$ appropriately. Another way to utilize this redundancy is by chosing $a$ such that the meson spectrum (at $\mathcal{O}\left(p^4\right)$) arises entirely from the leading terms in Eq.~\eqref{eq:LOChiralLag}. For example, denoting the up and down type masses in this gauge by $\mu_u$ and $\mu_d$ respectively, and taking the quark masses to be real which we will assume from now on, one gets 
\begin{equation}
    m_\pi^2 = B_0 (\mu_u+\mu_d) 
\end{equation}
in this gauge even from the $\mathcal{O}\left(p^4\right)$ Lagrangian.

An essential consequence of the above ambiguity is that fundamental/ultraviolet (electroweak generated) quark mass parameters of the Standard Model (SM) Lagrangian can  be extracted from the measured pion spectrum {\it only} up to an arbitrary shift~\cite{Kaplan:1986ru,Banks:1994yg}. Therefore, one needs to find an indirect route in order to check whether the observed meson spectrum is consistent with a zero up quark mass ansatz or not. The key is to first work in a gauge where the $\mathcal{O}\left(p^2\right)$ Lagrangian in Eq.~\eqref{eq:LOChiralLag} is matched to the quark Lagrangian of the SM (with mass parameters $m_{u,d,s}$). Then at $\mathcal{O}\left(p^4\right)$ one finds
\beq \label{eq:pionMassFull}
m_\pi^2 = \beta_1 (m_u + m_d)  + \beta_2 m_s (m_u + m_d) + \mathcal O(m_{u,d}^2),
\eeq
where 
\beq
\beta_1 = B_0, \quad \beta_2 = 16\frac{B_0^2}{F_0^2} (2 L_6 - L_4).
\eeq
This allows one to directly compare the $m_{u,d,s}$ parameters with the effective $\mu_{u,d,s}$ parameters since the pion masses are gauge invariant. In fact, it rather simplifies if the SM Lagrangian is characterized by a massless up quark, i.e. $m_u=0$, and if one further argues that $\mu_d \approx m_d$ and $\mu_s \approx m_s$~\cite{Dine:2014dga}
\beq
 \frac{\mu_u}{\mu_d} \simeq \frac{\beta_2}{\beta_1} \mu_s = 16 \frac{B_0 \mu_s}{F_0^2} (2 L_6 - L_4) \,.
 \label{eq:muudL}
\eeq 

Note that the value $\mu_u / \mu_d \simeq 0.5$ corresponds to $\beta_2 / \beta_1 \simeq 5~\mathrm{GeV}^{-1}$, a value that is commonly cited in the literature.
While lattice simulation results~\cite{FlavourLatticeAveragingGroup:2019iem,Alexandrou:2020bkd} report a highly constrained value of $\mu_u / \mu_d \lesssim 0.1$, unconstrained \chpt~fits to experimental data~\cite{Gasser:1984gg,Bijnens:2011tb,Bijnens:2014lea} prefer larger values of $L_4$ and $L_6$, allowing $\mu_u / \mu_d \simeq 0.5$ within $1\sigma$ uncertainty.

\section{The scalar potential in AMSB QCD}
The low-energy constants $L_i$ in the chiral Lagrangian, as well as $B_0$ and $F_0$, cannot be computed from first principles within QCD, since QCD gets strongly coupled at low energies. Determinations of the $L_i$ typically rely on phenomenological fits~\cite{Gasser:1984gg,Bijnens:2011tb,Bijnens:2014lea} or lattice simulations~\cite{Fodor:2016bgu,FlavourLatticeAveragingGroupFLAG:2024oxs}. Their leading order scaling can also be estimated as an expansion in the number of colors, in the large-$N$ limit~\cite{Gasser:1984gg,Leutwyler:1996sa,Kaiser:2000gs,Davies:2022ueb}. While it is generally believed that an extrapolation of large-$N$ results to $N=3$ gives a good qualitative description of QCD, conclusions derived from large-$N$ QCD should always be taken with a grain of salt.

Here we consider an alternative setup based on supersymmetric QCD (SQCD)~\cite{Affleck:1983mk,Seiberg:1994bz,Seiberg:1994pq} (see~\cite{Intriligator:1995au} for a review), which allows us to derive the chiral Lagrangian from first principles for any number of colors $N$, including $N=3$. We additionally assume that supersymmetry is softly broken via anomaly mediated supersymmetry breaking (AMSB)~\cite{Randall:1998uk,Giudice:1998xp}. AMSB ensures that SQCD has a QCD-like, i.e. chiral symmetry-breaking vacuum for $F < 3N$~\cite{Csaki:2022cyg}. Supersymmetry breaking lifts the masses of the superpartners and ensures that the pseudo-Nambu-Goldstone Bosons (pNGBs) from chiral symmetry breaking, i.e. the mesons, are the lightest states in the theory, just like in ordinary QCD. By explicitly integrating out heavier states in the scalar sector we can derive the chiral Lagrangian to any order in quark masses. However, note that, while this theory has many similarities with QCD, it is only QCD-like. The superpartners, and the gluinos in particular, still play an important role in the confinement dynamics, and only in the limit in which the supersymmetry breaking scale $m$ gets larger than the confinement scale $\Lambda$, one would recover ordinary QCD. However, when $m\sim \Lambda$ perturbativity is lost and this limit is not straightforward to take.

Our setup is a supersymmetric $SU(N)$ gauge theory with $F$ flavors of chiral superfield pairs in the fundamental and antifundamental representation of the gauge group. The effects of AMSB are incorporated using the conformal compensator formalism~\cite{Pomarol:1999ie}, where the supersymmetry breaking scale $m$ is the $F$-term of a chiral superfield that appears wherever scale invariance is broken. At tree-level it generates a contribution to the scalar potential which is given by~\cite{Csaki:2022cyg}
\bealg
V_\mathrm{tree} =& \pd_i W g^{ij^*} \pd_{j^*} W + m^* m \bigg( \pd_i K g^{ij^*} \pd_{j^*}K - K \bigg) \\
& + m \bigg( \pd_i W g^{ij^*}\pd_{j^*}K - 3 W \bigg) + c.c.,
\label{eq:vtree}
\eealg
where $g^{ij^*}$ is the inverse of the Kähler metric $g_{ij^*} = \pd_i\pd_{j^*} K$. 
Given our specific interest in $F=3$, this paper focuses on the cases where $F<N$ and $F=N$. The K\"ahler potential and superpotential of interest are given by
\bealg
W^{F<N} &= (N - F) \left( \frac{\Lambda^{3N-F}}{\det M} \right)^{1/(N-F)} + \tr[M_Q M], \\
K^{F<N} &= 2\,\tr\left[ \sqrt{M M^\dagger} \right]
\label{eq:kfln}
\eealg
for $F < N$~\cite{Davis:1983mz,Affleck:1983mk} and 
\bealg
W^{F=N} &= X\left( \frac{\det M - B \bar B}{\Lambda^{2N}} - 1 \right) + \tr[M_Q M], \\
K^{F=N} &= \frac{\tr[M M^\dagger]}{\alpha |\Lambda|^2} + \frac{X X^\dagger}{\beta |\Lambda|^4} + \frac{B B^\dagger}{\gamma|\Lambda|^{2N-2}} + \frac{\bar B \bar B^\dagger}{\delta|\Lambda|^{2N-2}}
\label{eq:kfen}
\eealg
for $F = N$~\cite{Seiberg:1994bz}. Here $M_{f f'} = \bar{Q}_f Q_{f'}$, $f,f'=1,\ldots ,F$ is the meson superfield, constructed from quark chiral superfields $Q_f, \bar{Q}_f$. $M_Q$ is the supersymmetric mass matrix for the quark superfields. $B, \bar{B}$ are completely antisymmetric color singlet combinations of the quark superfields $Q$ and $\bar{Q}$, i.e. $B = \epsilon^{f_1 \cdots f_F} \epsilon_{a_1 \cdots a_N} Q_{f_1}^{a_1} \cdots Q_{f_F}^{a_N}$, where $a_i$ are color indices.
$X$ corresponds to a Lagrange multiplier that implement the quantum modified constraint $\det M - B \bar{B}  = \Lambda^{2N}$ in the superpotential and $\alpha,\beta,\gamma,\delta$ 
are incalculable numbers.

Note that truncating the Kähler potential to only quadratic terms for $F=N$ is valid as long as $M /\Lambda^2, X/\Lambda^3, B/\Lambda^N, \bar{B}/\Lambda^N \ll 1$, which is not the case. This implies that there are higher-order terms in the Kähler potential which are as important as the leading order term. Our assumption is therefore that we still capture the qualitative behavior of the theory even if we neglect the higher-order terms. A more rigorous approach would be to start with $F=N+1$, which is under perturbative control, and introduce a heavy supersymmetric mass $\mu$ for one flavor, satisfying $\Lambda > \mu > m > M_Q$, and then integrate it out. In this framework, the Lagrange multiplier field $X$ naturally corresponds to the $M_{N+1,N+1}$ component of the meson field, providing a justification for the assumed form of its Kähler potential. It has been verified that this approach yields results consistent with our simplified method. 

Using the explicit expressions for $K$ and $W$, we obtain the scalar potential from~\Eq{eq:vtree} for each theory. Since we are interested in a QCD-like minimum and the chiral perturbation theory, we have $B = \bar B = 0$, and integrate out $X$ in the $F=N$ case. For simplicity, we also set $\theta=0$, i.e. $\arg\Lambda =0$ for the rest of the paper. For a derivation of the leading order chiral Lagrangian with $\theta\neq 0$, see for example~\cite{Dine:2016sgq,Csaki:2023yas,Csaki:2024lvk}.

\section{The Chiral Lagrangian in AMSB QCD}
As a next step we determine the chiral Lagrangian in AMSB QCD by integrating out the heavy degrees of freedom in the scalar sector at tree level. Note that the fermionic superpartners of the mesons contribute to the scalar potential only at the loop level. We will assume that $\Lambda \gg m \gg M_Q$ in order to maintain perturbative control. Following~\cite{Jungnickel:1997yu} we parameterize the meson matrix as 
\beq
M = S U, \quad S^\dagger = S, \quad {U}^\dagger {U} = \mathbf{I}_3.
\eeq
Hereafter, $M$ refers to the complex scalar component of the meson superfield.
Here ${U}$ contains the nine pNGBs in a nonlinear representation
\beq
U = \exp\left\{ -\frac{i}{3}\eta' \right\} \tilde U, \quad \det \tilde U = 1,
\eeq
where the $\eta'$ is the GB of the spontaneously broken $U(1)_A$, where we have absorbed the decay constant into the definiton of the field. Note that $U$ transforms as a bi-fundamental under $U(3)_L\times U(3)_R$, while $S$ is a singlet under $SU(3)_L$ and transforms in the adjoint representation of $SU(3)_R$. Therefore, $S$ contains nine real degrees of freedom (eight that transform in the adjoint representation of $SU(3)_R$ and one singlet).

To zeroth order in the quark masses $S$ obtains a vacuum expectation value (VEV) proportional to the identity, what spontaneously breaks $U(3)_L\times U(3)_R \rightarrow U(3)_V$, i.e.
\beq
\langle S\rangle = f_0^2\, \mathbf{I}_3 + \mathcal{O}(M_Q)\,,
\eeq
with
\bealg
f_0^{F < N} & = \left( \frac{N+3}{3N-3}\frac{\Lambda}{m}\right)^{\frac{N-3}{2N}} \Lambda \, , \\
f_0^{F = N} &= \Lambda - \frac{m^2}{6\alpha\beta \Lambda} + \mathcal O(m^4)\,.
\label{eq:f0}
\eealg
Plugging this into the scalar potential, one finds the leading-order chiral Lagrangian in Eq.~\eqref{eq:LOChiralLag} and identifies $F_0$ and $\chi = 2B_0 M_Q^\dagger$, where the explicit expression for $F_0$ and $B_0$ can be found in Table~\ref{tab:ls}.

In order to find higher-order terms in the chiral Lagrangian, we have to integrate out the radial modes in $S$, i.e. we have to perturbatively solve the equation of motion for $S$ as a function of $U$ to a given order in $M_Q$, i.e. we have to find $S=S[U]$, which minimizes the potential. In order to do so we parameterize $S$ as
\beq
S = f_0^2 \mathbf{I}_3 + S_1 + \cdots \equiv f_0^2 \mathbf{I}_3 + H + T \mathbf{I}_3 + \cdots,
\eeq
where $H = S_1 - \frac{1}{3} \tr[S_1] \mathbf{I}_3$ and $T =  \frac{1}{3} \tr[S_1]$. Note that $H$ and $T$ are of order $\mathcal O(M_Q)$ and $\mathcal O(\pd^2)$ by construction~\footnote{Note that minimizing the potential w.r.t. $S_1$ will also correct the VEV in a form that does not respect the chiral symmetry, i.e. $\langle S \rangle = f_0 \, \mathbf{I}_3 + a \, \tr[M_Q] \mathbf{I}_3 + b \, M_Q + \mathcal{O}(M_Q^2)$. This also happens in ordinary QCD~\cite{Gasser:1984gg}.}.
The relevant terms in the Lagrangian which include $H$ and $T$ take the form
\beq
\mathcal L^{(H,T)} = \frac{M_H^2}{2} \tr[H^2] - \tr[A^{(H)}[U] H] 
+ \frac{M_T^2}{2} T^2 - A^{(T)}[U] T,
\eeq
where $M_{(H,T)}$ is the effective mass of $H$ and $T$ and $A^{(H,T)}[U]$ is a $U$-dependent source term, which is different for each potential. $A^{(H,T)}$ can be decomposed into three parts $A_{M_Q}^{(H,T)}, A_\mathrm{kin}^{(H,T)}, A_{\eta'}^{(H,T)}$, each coming from the mass term, the kinetic term, and $\eta'$-dependent terms, respectively:
\bealg
A_{M_Q}^{(H)} &\propto \chic U + \Uc \chi - \frac{1}{3} \tr[\chic U + \Uc \chi], \\
A_{M_Q}^{(T)} &\propto \tr[\chic U + \Uc \chi], \\
A_{\mathrm{kin}}^{(H)} &\propto \pd_\mu U \pd^\mu \Uc - \frac{1}{3} \tr[\pd_\mu U \pd^\mu \Uc], \\
A_{\mathrm{kin}}^{(T)} &\propto \tr[\pd_\mu U \pd^\mu \Uc], \\
A_{\eta'}^{(H)} &= 0, \\
A_{\eta'}^{(T)} &\propto 1 - \cos\eta'.
\eealg
Note that at this order in $M_Q$ there is no mixing between $H$ and $T$.
Solving the equation of motion for $H$ and $T$ yields
\beq
H = \frac{A^{(H)}[U]}{M_H^2}, \quad T = \frac{A^{(T)}[U]}{M_T^2}.
\eeq
Substituting the solution back into $\mathcal L^{(H,T)}$ results in the effective Lagrangian of the nonlinear sigma model
\bealg
\mathcal L_1^{(H,T)} = & -\frac{1}{2M_H^2} \tr\left[\left(A^{(H)}_{M_Q} + A^{(H)}_\mathrm{kin} + A^{(H)}_{\eta'} \right)^2 \right] \\
& - \frac{1}{2M_T^2} \left(A^{(T)}_{M_Q} + A^{(T)}_\mathrm{kin} + A^{(T)}_{\eta'} \right)^2.
\label{eq:l1}
\eealg
By the same token, we integrate out $\eta'$, which contains the $\eta'$ meson:
\beq
\mathcal L_1^{(\eta')} = -\frac{1}{2M_\eta'^2} \left( A^{(\eta')}_{M_Q} + A^{(\eta')}_\mathrm{kin} \right)^2,
\eeq
where
\bealg
A_{M_Q}^{(\eta')} &\propto \tr[\chic U + \Uc \chi] \\
A_{\mathrm{kin}}^{(\eta')} & = 0.
\eealg
This term generates $\mathcal O_7$ which gives an additional contribution to the mass of the $\eta$ meson, due to $\eta-\eta'$ mixing.

\begin{table}[t]
\begin{tabular}{|c|c|c|}
\hline
Theories & $F=3<N$ & $F=N=3$ \\ \hline
$F_0$    & $2 f_0^{F<N} \sim \mathcal O(N^{1/2})$ & $ \frac{2\Lambda}{\sqrt\alpha} $ \\ \hline
$B_0$    & $\frac{7N-3}{N+3}\frac{m}{2}$ & $\frac{3\alpha m}{2}$ \\ \hline
$L_1$    & $-\frac{\Lambda^3}{36Nm^3} \sim \mathcal O(N^{0})$ & $ - \frac{\Lambda^2}{6\alpha m^2} + \frac{4}{9\alpha^2\beta}$ \\ \hline
$L_2$    & $0$ & $0$ \\ \hline
$L_3$    & $\frac{\Lambda^3}{108m^3} \sim \mathcal O(N^{1})$ & $ \frac{\Lambda^2}{2\alpha m^2} -\frac{1}{\alpha^2\beta}$ \\ \hline
$L_4$    & $-\frac{\Lambda^3}{42Nm^3} \sim \mathcal O(N^{0})$ & $- \frac{\Lambda^2}{18\alpha m^2} + \frac{16}{81\alpha^2\beta} $ \\ \hline
$L_5$    & $\frac{\Lambda^3}{378m^3} \sim \mathcal O(N^{1})$ & $\frac{\Lambda^2}{6\alpha m^2} - \frac{13}{27\alpha^2\beta} $ \\ \hline
$L_6$    & $-\frac{5\Lambda^3}{1764Nm^3} \sim \mathcal O(N^{0})$ & $\frac{\Lambda^2}{216\alpha m^2} + \frac{5}{486\alpha^2\beta}$ \\ \hline
$L_7$    & $-\frac{N\Lambda^3}{972m^3} \sim \mathcal O(N^{2})$ & $\frac{\Lambda^2}{36\alpha m^2} -\frac{19}{324\alpha^2\beta}$ \\ \hline
$L_8$    & $\frac{\Lambda^3}{5292m^3} \sim \mathcal O(N^{1})$ & $\frac{\Lambda^2}{72\alpha m^2} -\frac{17}{324\alpha^2\beta}$ \\ \hline
\end{tabular}
\caption{The parameters of \chpt~in the lowest and next-to-leading order quark mass expansion for $F < N$ and $F = N$. $F$ is specifically taken as 3. Large $N$ counting is added in the $F<N$ columnn.}
\label{tab:ls}
\end{table}
The low-energy constants obtained from our matching procedure  
for $F < N$ and $F = N$ are listed in Table~\ref{tab:ls}.
For $F < N$, they are expanded in powers of $1/N$, while for $F = N$, the expansion is in powers of $1/\beta$. Note that the limit $\beta \to \infty$ corresponds to $X$ becoming non-dynamical, i.e. an infinitely heavy genuine Lagrange multiplier. 
In this regime, the vacuum structure can still be determined even when $m \to \Lambda$, provided that $\beta \gg 1$.

The large-$N$ counting is added for the $F<N$ case, using the $N$-dependence in the translation from the holomorphic to the canonical scale $|\Lambda|\propto N^{1/3}\Lambda_c$~\cite{Dine:2016sgq,Csaki:2023yas}, where $\Lambda_c$ is the physical QCD scale
\beq
|\Lambda| = \left( \frac{b_0}{8\pi^2} \right)^{\frac{b_1}{2b_0^2}} \Lambda_c\,,
\eeq
with $b_0 = 3N-F$ and $b_1 = 6N^2 - 2N F - \frac{4 F(N^2-1)}{2N}$.
Note that the large $N$-behavior for the \chpt~parameters, except for $L_1$ and $L_2$, matches estimates in the literature~\cite{Gasser:1984gg,Leutwyler:1996sa,Kaiser:2000gs,Davies:2022ueb}. $L_1$ and $L_2$ are known to scale as $N^1$, and are typically dominated by the exchange of vector mesons~\cite{Ecker:1988te}. However, since the potentials considered in this paper include only pseudoscalar fields and omit vector and axial-vector resonances, the contributions to $L_1$ and $L_2$ are suppressed.

\section{Dynamical up quark mass in AMSB QCD}
We are now ready to present the dynamical up quark mass in QCD-like theories.  
Using \Eq{eq:muudL} along with the values of $L_4$, $L_6$, and $B_0$ from Table~\ref{tab:ls}, we find
\beq
\frac{\mu_u}{\mu_d} \simeq
\begin{cases}
 \displaystyle \frac{16}{147}\frac{(N-1)(7N-3)}{N(N+3)^2} \left( \frac{N+3}{3N-3} \frac{\Lambda}{m} \right)^{3/N} \frac{\mu_s}{m} \qquad \\ \hskip 15em (F=3<N)\\[2ex] \\
 \displaystyle \frac{7\alpha}{18}\frac{\mu_s}{m} \hskip 12.4em (F=N=3).\\[2ex] 
\end{cases}
\eeq
Note that due to our assumption $m > M_Q$, $\mu_u/\mu_d$ does not diverge when $m \to 0$ since $m_s$ must also vanish in this limit. We can see that these expressions are sizable, and in particular for the $F=N$ case $\mu_u/\mu_d = {\cal O}(1)$ as long as $\alpha  = {\cal O}(1)$ and $\mu_s/m$ is chosen to be not too small. 

It remains unclear whether or how these AMSB QCD results can be reliably extrapolated to real QCD in the limit where the SUSY breaking scale becomes large ($m\gg \Lambda$). One naive guess would be to set $m\sim {\cal O}(\Lambda )$, assuming that all states obtaining masses from SUSY breaking decouple from the dynamics once they are heavier than $\Lambda$. Note that for $F < N$, the theory allows for a self-consistent extrapolation to $m\sim {\cal O}(\Lambda )$, as the effective theory is fully solvable in this limit. This also holds for $F = N$ as long as $\beta \gg 1$, which is expected since $X$ acts as a Lagrange multiplier that enforces the quantum modified constraint on the moduli space. Using such a naive extrapolation one would obtain the following numerical values: 
\beq
\frac{\mu_u}{\mu_d} \simeq 0.01 \qquad
\left(\substack{F =3, N = 4 \\ \Lambda_c = m , \, \mu_s / m = 0.5}\right).
\eeq
For $F = N$ the result depends on the uncalculable parameter $\alpha$, usually assumed to be  $\alpha \sim \mathcal{O}(1)$. As an illustrative example, choosing $\alpha = 2$ leads to
\beq
\frac{\mu_u}{\mu_d} \simeq 0.4 \qquad 
\left(\substack{F = N = 3 \\  \alpha = 2 , \, \mu_s / m = 0.5}\right).
\eeq
We should again emphasize that these numerical values are obtained by extrapolating outside the regime of validity of our theory. 

The large-$N$ behavior of our result matches the expectation that $\mu_u/\mu_d$ is suppressed in this limit, vanishing altogether at leading order in $1/N$.
However, in the case of $F = N=3$ a sizable $\mu_u$ is generated. It is probably not too surprising that the result for this case is quite different (and much bigger) than for the $F<N$ case: the $F=N$ case is truly strongly coupled, and the origin of chiral symmetry breaking lies in the confining dynamics (while for $F<N$ the theory is actually weakly coupled).  
In fact the resulting dynamical contribution for the $F=N$ case could potentially account for the entire mass of the up quark without requiring an explicit up quark Yukawa coupling, thereby rendering the QCD $\theta$ angle unphysical.

The fundamental origin of the dynamical up-quark mass is strongly tied to the generation of the effective low-energy superpotential from which it arises, in combination with the supersymmetric mass term and the supersymmetry breaking effects. It is well-known that for $F=N-1$ the ADS superpotential is generated by weakly-coupled instantons~\cite{Affleck:1983mk}, while for $F<N-1$ it is due to gaugino condensation in the unbroken $SU(N-F)$. For $F=N$   a quantum modified constraint on the moduli space arises from the dynamics, which is implemented by the Lagrange multiplier field $X$. The exact origin of the quantum modified constraint remains unknown, however, it is consistent with being a strongly-coupled instanton effect.
 
\section{Acknowledgments}

\begin{acknowledgments}
We thank  Brando Bellazzini, Michael Dine, Sungwoo Hong, Ryuichiro Kitano, Hitoshi Murayama and Bea Noether for useful discussions and feedbacks. 
CC and TY are supported in part by the NSF grant PHY-2309456. CC is supported in part by the US-Israeli BSF grant 2016153. TY is funded by the Samsung Science and Technology Foundation under Project Number SSTF-BA2201-06. MR is supported by NSF Grant PHY-2310429, Simons Investigator Award No.~824870, DOE HEP QuantISED award \#100495, the Gordon and Betty Moore Foundation Grant GBMF7946, and the U.S.~Department of Energy (DOE), Office of Science, National Quantum Information Science Research Centers, Superconducting Quantum Materials and Systems Center (SQMS) under contract No.~DEAC02-07CH11359. TR was supported in part by the Department of Atomic Energy, Government of India, under Project Identification Number RTI 4002. 

\end{acknowledgments}

\bibliography{bibtex}

\begin{thebibliography}{31}%
\makeatletter
\providecommand \@ifxundefined [1]{%
 \@ifx{#1\undefined}
}%
\providecommand \@ifnum [1]{%
 \ifnum #1\expandafter \@firstoftwo
 \else \expandafter \@secondoftwo
 \fi
}%
\providecommand \@ifx [1]{%
 \ifx #1\expandafter \@firstoftwo
 \else \expandafter \@secondoftwo
 \fi
}%
\providecommand \natexlab [1]{#1}%
\providecommand \enquote  [1]{``#1''}%
\providecommand \bibnamefont  [1]{#1}%
\providecommand \bibfnamefont [1]{#1}%
\providecommand \citenamefont [1]{#1}%
\providecommand \href@noop [0]{\@secondoftwo}%
\providecommand \href [0]{\begingroup \@sanitize@url \@href}%
\providecommand \@href[1]{\@@startlink{#1}\@@href}%
\providecommand \@@href[1]{\endgroup#1\@@endlink}%
\providecommand \@sanitize@url [0]{\catcode `\\12\catcode `\$12\catcode
  `\&12\catcode `\#12\catcode `\^12\catcode `\_12\catcode `\%12\relax}%
\providecommand \@@startlink[1]{}%
\providecommand \@@endlink[0]{}%
\providecommand \url  [0]{\begingroup\@sanitize@url \@url }%
\providecommand \@url [1]{\endgroup\@href {#1}{\urlprefix }}%
\providecommand \urlprefix  [0]{URL }%
\providecommand \Eprint [0]{\href }%
\providecommand \doibase [0]{https://doi.org/}%
\providecommand \selectlanguage [0]{\@gobble}%
\providecommand \bibinfo  [0]{\@secondoftwo}%
\providecommand \bibfield  [0]{\@secondoftwo}%
\providecommand \translation [1]{[#1]}%
\providecommand \BibitemOpen [0]{}%
\providecommand \bibitemStop [0]{}%
\providecommand \bibitemNoStop [0]{.\EOS\space}%
\providecommand \EOS [0]{\spacefactor3000\relax}%
\providecommand \BibitemShut  [1]{\csname bibitem#1\endcsname}%
\let\auto@bib@innerbib\@empty
\bibitem [{\citenamefont {Kaplan}\ and\ \citenamefont
  {Manohar}(1986)}]{Kaplan:1986ru}%
  \BibitemOpen
  \bibfield  {author} {\bibinfo {author} {\bibfnamefont {D.~B.}\ \bibnamefont
  {Kaplan}}\ and\ \bibinfo {author} {\bibfnamefont {A.~V.}\ \bibnamefont
  {Manohar}},\ }\bibfield  {title} {\bibinfo {title} {{Current Mass Ratios of
  the Light Quarks}},\ }\href {https://doi.org/10.1103/PhysRevLett.56.2004}
  {\bibfield  {journal} {\bibinfo  {journal} {Phys. Rev. Lett.}\ }\textbf
  {\bibinfo {volume} {56}},\ \bibinfo {pages} {2004} (\bibinfo {year}
  {1986})}\BibitemShut {NoStop}%
\bibitem [{\citenamefont {Cs\'aki}\ \emph
  {et~al.}(2024{\natexlab{a}})\citenamefont {Cs\'aki}, \citenamefont
  {D'Agnolo}, \citenamefont {Kuflik},\ and\ \citenamefont
  {Ruhdorfer}}]{Csaki:2023ziz}%
  \BibitemOpen
  \bibfield  {author} {\bibinfo {author} {\bibfnamefont {C.}~\bibnamefont
  {Cs\'aki}}, \bibinfo {author} {\bibfnamefont {R.~T.}\ \bibnamefont
  {D'Agnolo}}, \bibinfo {author} {\bibfnamefont {E.}~\bibnamefont {Kuflik}},\
  and\ \bibinfo {author} {\bibfnamefont {M.}~\bibnamefont {Ruhdorfer}},\
  }\bibfield  {title} {\bibinfo {title} {{Instanton NDA and applications to
  axion models}},\ }\href {https://doi.org/10.1007/JHEP04(2024)074} {\bibfield
  {journal} {\bibinfo  {journal} {JHEP}\ }\textbf {\bibinfo {volume} {04}},\
  \bibinfo {pages} {074}},\ \Eprint {https://arxiv.org/abs/2311.09285}
  {arXiv:2311.09285 [hep-ph]} \BibitemShut {NoStop}%
\bibitem [{\citenamefont {Fodor}\ \emph {et~al.}(2016)\citenamefont {Fodor},
  \citenamefont {Hoelbling}, \citenamefont {Krieg}, \citenamefont {Lellouch},
  \citenamefont {Lippert}, \citenamefont {Portelli}, \citenamefont {Sastre},
  \citenamefont {Szabo},\ and\ \citenamefont {Varnhorst}}]{Fodor:2016bgu}%
  \BibitemOpen
  \bibfield  {author} {\bibinfo {author} {\bibfnamefont {Z.}~\bibnamefont
  {Fodor}}, \bibinfo {author} {\bibfnamefont {C.}~\bibnamefont {Hoelbling}},
  \bibinfo {author} {\bibfnamefont {S.}~\bibnamefont {Krieg}}, \bibinfo
  {author} {\bibfnamefont {L.}~\bibnamefont {Lellouch}}, \bibinfo {author}
  {\bibfnamefont {T.}~\bibnamefont {Lippert}}, \bibinfo {author} {\bibfnamefont
  {A.}~\bibnamefont {Portelli}}, \bibinfo {author} {\bibfnamefont
  {A.}~\bibnamefont {Sastre}}, \bibinfo {author} {\bibfnamefont {K.~K.}\
  \bibnamefont {Szabo}},\ and\ \bibinfo {author} {\bibfnamefont
  {L.}~\bibnamefont {Varnhorst}},\ }\bibfield  {title} {\bibinfo {title} {{Up
  and down quark masses and corrections to Dashen's theorem from lattice QCD
  and quenched QED}},\ }\href {https://doi.org/10.1103/PhysRevLett.117.082001}
  {\bibfield  {journal} {\bibinfo  {journal} {Phys. Rev. Lett.}\ }\textbf
  {\bibinfo {volume} {117}},\ \bibinfo {pages} {082001} (\bibinfo {year}
  {2016})},\ \Eprint {https://arxiv.org/abs/1604.07112} {arXiv:1604.07112
  [hep-lat]} \BibitemShut {NoStop}%
\bibitem [{\citenamefont {Aoki}\ \emph {et~al.}(2024)\citenamefont {Aoki} \emph
  {et~al.}}]{FlavourLatticeAveragingGroupFLAG:2024oxs}%
  \BibitemOpen
  \bibfield  {author} {\bibinfo {author} {\bibfnamefont {Y.}~\bibnamefont
  {Aoki}} \emph {et~al.} (\bibinfo {collaboration} {Flavour Lattice Averaging
  Group (FLAG)}),\ }\href@noop {} {\bibinfo {title} {{FLAG Review 2024}}}
  (\bibinfo {year} {2024}),\ \Eprint {https://arxiv.org/abs/2411.04268}
  {arXiv:2411.04268 [hep-lat]} \BibitemShut {NoStop}%
\bibitem [{\citenamefont {Frison}\ \emph {et~al.}(2016)\citenamefont {Frison},
  \citenamefont {Kitano},\ and\ \citenamefont {Yamada}}]{Frison:2016rnq}%
  \BibitemOpen
  \bibfield  {author} {\bibinfo {author} {\bibfnamefont {J.}~\bibnamefont
  {Frison}}, \bibinfo {author} {\bibfnamefont {R.}~\bibnamefont {Kitano}},\
  and\ \bibinfo {author} {\bibfnamefont {N.}~\bibnamefont {Yamada}},\
  }\bibfield  {title} {\bibinfo {title} {{$N_f=1+2$ mass dependence of the
  topological susceptibility}},\ }\href {https://doi.org/10.22323/1.256.0323}
  {\bibfield  {journal} {\bibinfo  {journal} {PoS}\ }\textbf {\bibinfo {volume}
  {LATTICE2016}},\ \bibinfo {pages} {323} (\bibinfo {year} {2016})},\ \Eprint
  {https://arxiv.org/abs/1611.07150} {arXiv:1611.07150 [hep-lat]} \BibitemShut
  {NoStop}%
\bibitem [{\citenamefont {Gasser}\ and\ \citenamefont
  {Leutwyler}(1985)}]{Gasser:1984gg}%
  \BibitemOpen
  \bibfield  {author} {\bibinfo {author} {\bibfnamefont {J.}~\bibnamefont
  {Gasser}}\ and\ \bibinfo {author} {\bibfnamefont {H.}~\bibnamefont
  {Leutwyler}},\ }\bibfield  {title} {\bibinfo {title} {{Chiral Perturbation
  Theory: Expansions in the Mass of the Strange Quark}},\ }\href
  {https://doi.org/10.1016/0550-3213(85)90492-4} {\bibfield  {journal}
  {\bibinfo  {journal} {Nucl. Phys. B}\ }\textbf {\bibinfo {volume} {250}},\
  \bibinfo {pages} {465} (\bibinfo {year} {1985})}\BibitemShut {NoStop}%
\bibitem [{\citenamefont {Bijnens}\ and\ \citenamefont
  {Jemos}(2012)}]{Bijnens:2011tb}%
  \BibitemOpen
  \bibfield  {author} {\bibinfo {author} {\bibfnamefont {J.}~\bibnamefont
  {Bijnens}}\ and\ \bibinfo {author} {\bibfnamefont {I.}~\bibnamefont
  {Jemos}},\ }\bibfield  {title} {\bibinfo {title} {{A new global fit of the
  $L^r_i$ at next-to-next-to-leading order in Chiral Perturbation Theory}},\
  }\href {https://doi.org/10.1016/j.nuclphysb.2011.09.013} {\bibfield
  {journal} {\bibinfo  {journal} {Nucl. Phys. B}\ }\textbf {\bibinfo {volume}
  {854}},\ \bibinfo {pages} {631} (\bibinfo {year} {2012})},\ \Eprint
  {https://arxiv.org/abs/1103.5945} {arXiv:1103.5945 [hep-ph]} \BibitemShut
  {NoStop}%
\bibitem [{\citenamefont {Bijnens}\ and\ \citenamefont
  {Ecker}(2014)}]{Bijnens:2014lea}%
  \BibitemOpen
  \bibfield  {author} {\bibinfo {author} {\bibfnamefont {J.}~\bibnamefont
  {Bijnens}}\ and\ \bibinfo {author} {\bibfnamefont {G.}~\bibnamefont
  {Ecker}},\ }\bibfield  {title} {\bibinfo {title} {{Mesonic low-energy
  constants}},\ }\href {https://doi.org/10.1146/annurev-nucl-102313-025528}
  {\bibfield  {journal} {\bibinfo  {journal} {Ann. Rev. Nucl. Part. Sci.}\
  }\textbf {\bibinfo {volume} {64}},\ \bibinfo {pages} {149} (\bibinfo {year}
  {2014})},\ \Eprint {https://arxiv.org/abs/1405.6488} {arXiv:1405.6488
  [hep-ph]} \BibitemShut {NoStop}%
\bibitem [{\citenamefont {Dine}\ \emph {et~al.}(2015)\citenamefont {Dine},
  \citenamefont {Draper},\ and\ \citenamefont {Festuccia}}]{Dine:2014dga}%
  \BibitemOpen
  \bibfield  {author} {\bibinfo {author} {\bibfnamefont {M.}~\bibnamefont
  {Dine}}, \bibinfo {author} {\bibfnamefont {P.}~\bibnamefont {Draper}},\ and\
  \bibinfo {author} {\bibfnamefont {G.}~\bibnamefont {Festuccia}},\ }\bibfield
  {title} {\bibinfo {title} {{Instanton Effects in Three Flavor QCD}},\ }\href
  {https://doi.org/10.1103/PhysRevD.92.054004} {\bibfield  {journal} {\bibinfo
  {journal} {Phys. Rev. D}\ }\textbf {\bibinfo {volume} {92}},\ \bibinfo
  {pages} {054004} (\bibinfo {year} {2015})},\ \Eprint
  {https://arxiv.org/abs/1410.8505} {arXiv:1410.8505 [hep-ph]} \BibitemShut
  {NoStop}%
\bibitem [{\citenamefont {Leutwyler}(1996)}]{Leutwyler:1996sa}%
  \BibitemOpen
  \bibfield  {author} {\bibinfo {author} {\bibfnamefont {H.}~\bibnamefont
  {Leutwyler}},\ }\bibfield  {title} {\bibinfo {title} {{Bounds on the light
  quark masses}},\ }\href {https://doi.org/10.1016/0370-2693(96)85876-X}
  {\bibfield  {journal} {\bibinfo  {journal} {Phys. Lett. B}\ }\textbf
  {\bibinfo {volume} {374}},\ \bibinfo {pages} {163} (\bibinfo {year}
  {1996})},\ \Eprint {https://arxiv.org/abs/hep-ph/9601234}
  {arXiv:hep-ph/9601234} \BibitemShut {NoStop}%
\bibitem [{\citenamefont {Kaiser}\ and\ \citenamefont
  {Leutwyler}(2000)}]{Kaiser:2000gs}%
  \BibitemOpen
  \bibfield  {author} {\bibinfo {author} {\bibfnamefont {R.}~\bibnamefont
  {Kaiser}}\ and\ \bibinfo {author} {\bibfnamefont {H.}~\bibnamefont
  {Leutwyler}},\ }\bibfield  {title} {\bibinfo {title} {{Large N(c) in chiral
  perturbation theory}},\ }\href {https://doi.org/10.1007/s100520000499}
  {\bibfield  {journal} {\bibinfo  {journal} {Eur. Phys. J. C}\ }\textbf
  {\bibinfo {volume} {17}},\ \bibinfo {pages} {623} (\bibinfo {year} {2000})},\
  \Eprint {https://arxiv.org/abs/hep-ph/0007101} {arXiv:hep-ph/0007101}
  \BibitemShut {NoStop}%
\bibitem [{\citenamefont {Davies}\ \emph {et~al.}(2022)\citenamefont {Davies},
  \citenamefont {Dine},\ and\ \citenamefont {Lehmann}}]{Davies:2022ueb}%
  \BibitemOpen
  \bibfield  {author} {\bibinfo {author} {\bibfnamefont {D.}~\bibnamefont
  {Davies}}, \bibinfo {author} {\bibfnamefont {M.}~\bibnamefont {Dine}},\ and\
  \bibinfo {author} {\bibfnamefont {B.~V.}\ \bibnamefont {Lehmann}},\
  }\href@noop {} {\bibinfo {title} {{Light Quarks at Large $N$}}} (\bibinfo
  {year} {2022}),\ \Eprint {https://arxiv.org/abs/2201.05719} {arXiv:2201.05719
  [hep-ph]} \BibitemShut {NoStop}%
\bibitem [{\citenamefont {Murayama}(2021)}]{Murayama:2021xfj}%
  \BibitemOpen
  \bibfield  {author} {\bibinfo {author} {\bibfnamefont {H.}~\bibnamefont
  {Murayama}},\ }\bibfield  {title} {\bibinfo {title} {{Some Exact Results in
  QCD-like Theories}},\ }\href {https://doi.org/10.1103/PhysRevLett.126.251601}
  {\bibfield  {journal} {\bibinfo  {journal} {Phys. Rev. Lett.}\ }\textbf
  {\bibinfo {volume} {126}},\ \bibinfo {pages} {251601} (\bibinfo {year}
  {2021})},\ \Eprint {https://arxiv.org/abs/2104.01179} {arXiv:2104.01179
  [hep-th]} \BibitemShut {NoStop}%
\bibitem [{\citenamefont {Cs\'aki}\ \emph
  {et~al.}(2023{\natexlab{a}})\citenamefont {Cs\'aki}, \citenamefont {Gomes},
  \citenamefont {Murayama}, \citenamefont {Noether}, \citenamefont {Varier},\
  and\ \citenamefont {Telem}}]{Csaki:2022cyg}%
  \BibitemOpen
  \bibfield  {author} {\bibinfo {author} {\bibfnamefont {C.}~\bibnamefont
  {Cs\'aki}}, \bibinfo {author} {\bibfnamefont {A.}~\bibnamefont {Gomes}},
  \bibinfo {author} {\bibfnamefont {H.}~\bibnamefont {Murayama}}, \bibinfo
  {author} {\bibfnamefont {B.}~\bibnamefont {Noether}}, \bibinfo {author}
  {\bibfnamefont {D.~R.}\ \bibnamefont {Varier}},\ and\ \bibinfo {author}
  {\bibfnamefont {O.}~\bibnamefont {Telem}},\ }\bibfield  {title} {\bibinfo
  {title} {{Guide to anomaly-mediated supersymmetry-breaking QCD}},\ }\href
  {https://doi.org/10.1103/PhysRevD.107.054015} {\bibfield  {journal} {\bibinfo
   {journal} {Phys. Rev. D}\ }\textbf {\bibinfo {volume} {107}},\ \bibinfo
  {pages} {054015} (\bibinfo {year} {2023}{\natexlab{a}})},\ \Eprint
  {https://arxiv.org/abs/2212.03260} {arXiv:2212.03260 [hep-th]} \BibitemShut
  {NoStop}%
\bibitem [{\citenamefont {Gell-Mann}\ \emph {et~al.}(1968)\citenamefont
  {Gell-Mann}, \citenamefont {Oakes},\ and\ \citenamefont
  {Renner}}]{Gell-Mann:1968hlm}%
  \BibitemOpen
  \bibfield  {author} {\bibinfo {author} {\bibfnamefont {M.}~\bibnamefont
  {Gell-Mann}}, \bibinfo {author} {\bibfnamefont {R.~J.}\ \bibnamefont
  {Oakes}},\ and\ \bibinfo {author} {\bibfnamefont {B.}~\bibnamefont
  {Renner}},\ }\bibfield  {title} {\bibinfo {title} {{Behavior of current
  divergences under SU(3) x SU(3)}},\ }\href
  {https://doi.org/10.1103/PhysRev.175.2195} {\bibfield  {journal} {\bibinfo
  {journal} {Phys. Rev.}\ }\textbf {\bibinfo {volume} {175}},\ \bibinfo {pages}
  {2195} (\bibinfo {year} {1968})}\BibitemShut {NoStop}%
\bibitem [{\citenamefont {Banks}\ \emph {et~al.}(1994)\citenamefont {Banks},
  \citenamefont {Nir},\ and\ \citenamefont {Seiberg}}]{Banks:1994yg}%
  \BibitemOpen
  \bibfield  {author} {\bibinfo {author} {\bibfnamefont {T.}~\bibnamefont
  {Banks}}, \bibinfo {author} {\bibfnamefont {Y.}~\bibnamefont {Nir}},\ and\
  \bibinfo {author} {\bibfnamefont {N.}~\bibnamefont {Seiberg}},\ }\bibfield
  {title} {\bibinfo {title} {{Missing (up) mass, accidental anomalous
  symmetries, and the strong CP problem}},\ }in\ \href@noop {} {\emph {\bibinfo
  {booktitle} {{2nd IFT Workshop on Yukawa Couplings and the Origins of
  Mass}}}}\ (\bibinfo {year} {1994})\ pp.\ \bibinfo {pages} {26--41},\ \Eprint
  {https://arxiv.org/abs/hep-ph/9403203} {arXiv:hep-ph/9403203} \BibitemShut
  {NoStop}%
\bibitem [{\citenamefont {Aoki}\ \emph {et~al.}(2020)\citenamefont {Aoki} \emph
  {et~al.}}]{FlavourLatticeAveragingGroup:2019iem}%
  \BibitemOpen
  \bibfield  {author} {\bibinfo {author} {\bibfnamefont {S.}~\bibnamefont
  {Aoki}} \emph {et~al.} (\bibinfo {collaboration} {Flavour Lattice Averaging
  Group}),\ }\bibfield  {title} {\bibinfo {title} {{FLAG Review 2019: Flavour
  Lattice Averaging Group (FLAG)}},\ }\href
  {https://doi.org/10.1140/epjc/s10052-019-7354-7} {\bibfield  {journal}
  {\bibinfo  {journal} {Eur. Phys. J. C}\ }\textbf {\bibinfo {volume} {80}},\
  \bibinfo {pages} {113} (\bibinfo {year} {2020})},\ \Eprint
  {https://arxiv.org/abs/1902.08191} {arXiv:1902.08191 [hep-lat]} \BibitemShut
  {NoStop}%
\bibitem [{\citenamefont {Alexandrou}\ \emph {et~al.}(2020)\citenamefont
  {Alexandrou}, \citenamefont {Finkenrath}, \citenamefont {Funcke},
  \citenamefont {Jansen}, \citenamefont {Kostrzewa}, \citenamefont {Pittler},\
  and\ \citenamefont {Urbach}}]{Alexandrou:2020bkd}%
  \BibitemOpen
  \bibfield  {author} {\bibinfo {author} {\bibfnamefont {C.}~\bibnamefont
  {Alexandrou}}, \bibinfo {author} {\bibfnamefont {J.}~\bibnamefont
  {Finkenrath}}, \bibinfo {author} {\bibfnamefont {L.}~\bibnamefont {Funcke}},
  \bibinfo {author} {\bibfnamefont {K.}~\bibnamefont {Jansen}}, \bibinfo
  {author} {\bibfnamefont {B.}~\bibnamefont {Kostrzewa}}, \bibinfo {author}
  {\bibfnamefont {F.}~\bibnamefont {Pittler}},\ and\ \bibinfo {author}
  {\bibfnamefont {C.}~\bibnamefont {Urbach}},\ }\bibfield  {title} {\bibinfo
  {title} {{Ruling Out the Massless Up-Quark Solution to the Strong $\pmb{CP}$
  Problem by Computing the Topological Mass Contribution with Lattice QCD}},\
  }\href {https://doi.org/10.1103/PhysRevLett.125.232001} {\bibfield  {journal}
  {\bibinfo  {journal} {Phys. Rev. Lett.}\ }\textbf {\bibinfo {volume} {125}},\
  \bibinfo {pages} {232001} (\bibinfo {year} {2020})},\ \Eprint
  {https://arxiv.org/abs/2002.07802} {arXiv:2002.07802 [hep-lat]} \BibitemShut
  {NoStop}%
\bibitem [{\citenamefont {Affleck}\ \emph {et~al.}(1984)\citenamefont
  {Affleck}, \citenamefont {Dine},\ and\ \citenamefont
  {Seiberg}}]{Affleck:1983mk}%
  \BibitemOpen
  \bibfield  {author} {\bibinfo {author} {\bibfnamefont {I.}~\bibnamefont
  {Affleck}}, \bibinfo {author} {\bibfnamefont {M.}~\bibnamefont {Dine}},\ and\
  \bibinfo {author} {\bibfnamefont {N.}~\bibnamefont {Seiberg}},\ }\bibfield
  {title} {\bibinfo {title} {{Dynamical Supersymmetry Breaking in
  Supersymmetric QCD}},\ }\href {https://doi.org/10.1016/0550-3213(84)90058-0}
  {\bibfield  {journal} {\bibinfo  {journal} {Nucl. Phys. B}\ }\textbf
  {\bibinfo {volume} {241}},\ \bibinfo {pages} {493} (\bibinfo {year}
  {1984})}\BibitemShut {NoStop}%
\bibitem [{\citenamefont {Seiberg}(1994)}]{Seiberg:1994bz}%
  \BibitemOpen
  \bibfield  {author} {\bibinfo {author} {\bibfnamefont {N.}~\bibnamefont
  {Seiberg}},\ }\bibfield  {title} {\bibinfo {title} {{Exact results on the
  space of vacua of four-dimensional SUSY gauge theories}},\ }\href
  {https://doi.org/10.1103/PhysRevD.49.6857} {\bibfield  {journal} {\bibinfo
  {journal} {Phys. Rev. D}\ }\textbf {\bibinfo {volume} {49}},\ \bibinfo
  {pages} {6857} (\bibinfo {year} {1994})},\ \Eprint
  {https://arxiv.org/abs/hep-th/9402044} {arXiv:hep-th/9402044} \BibitemShut
  {NoStop}%
\bibitem [{\citenamefont {Seiberg}(1995)}]{Seiberg:1994pq}%
  \BibitemOpen
  \bibfield  {author} {\bibinfo {author} {\bibfnamefont {N.}~\bibnamefont
  {Seiberg}},\ }\bibfield  {title} {\bibinfo {title} {{Electric - magnetic
  duality in supersymmetric nonAbelian gauge theories}},\ }\href
  {https://doi.org/10.1016/0550-3213(94)00023-8} {\bibfield  {journal}
  {\bibinfo  {journal} {Nucl. Phys. B}\ }\textbf {\bibinfo {volume} {435}},\
  \bibinfo {pages} {129} (\bibinfo {year} {1995})},\ \Eprint
  {https://arxiv.org/abs/hep-th/9411149} {arXiv:hep-th/9411149} \BibitemShut
  {NoStop}%
\bibitem [{\citenamefont {Intriligator}\ and\ \citenamefont
  {Seiberg}(1996)}]{Intriligator:1995au}%
  \BibitemOpen
  \bibfield  {author} {\bibinfo {author} {\bibfnamefont {K.~A.}\ \bibnamefont
  {Intriligator}}\ and\ \bibinfo {author} {\bibfnamefont {N.}~\bibnamefont
  {Seiberg}},\ }\bibfield  {title} {\bibinfo {title} {{Lectures on
  supersymmetric gauge theories and electric-magnetic duality}},\ }\href
  {https://doi.org/10.1016/0920-5632(95)00626-5} {\bibfield  {journal}
  {\bibinfo  {journal} {Nucl. Phys. B Proc. Suppl.}\ }\textbf {\bibinfo
  {volume} {45BC}},\ \bibinfo {pages} {1} (\bibinfo {year} {1996})},\ \Eprint
  {https://arxiv.org/abs/hep-th/9509066} {arXiv:hep-th/9509066} \BibitemShut
  {NoStop}%
\bibitem [{\citenamefont {Randall}\ and\ \citenamefont
  {Sundrum}(1999)}]{Randall:1998uk}%
  \BibitemOpen
  \bibfield  {author} {\bibinfo {author} {\bibfnamefont {L.}~\bibnamefont
  {Randall}}\ and\ \bibinfo {author} {\bibfnamefont {R.}~\bibnamefont
  {Sundrum}},\ }\bibfield  {title} {\bibinfo {title} {{Out of this world
  supersymmetry breaking}},\ }\href
  {https://doi.org/10.1016/S0550-3213(99)00359-4} {\bibfield  {journal}
  {\bibinfo  {journal} {Nucl. Phys. B}\ }\textbf {\bibinfo {volume} {557}},\
  \bibinfo {pages} {79} (\bibinfo {year} {1999})},\ \Eprint
  {https://arxiv.org/abs/hep-th/9810155} {arXiv:hep-th/9810155} \BibitemShut
  {NoStop}%
\bibitem [{\citenamefont {Giudice}\ \emph {et~al.}(1998)\citenamefont
  {Giudice}, \citenamefont {Luty}, \citenamefont {Murayama},\ and\
  \citenamefont {Rattazzi}}]{Giudice:1998xp}%
  \BibitemOpen
  \bibfield  {author} {\bibinfo {author} {\bibfnamefont {G.~F.}\ \bibnamefont
  {Giudice}}, \bibinfo {author} {\bibfnamefont {M.~A.}\ \bibnamefont {Luty}},
  \bibinfo {author} {\bibfnamefont {H.}~\bibnamefont {Murayama}},\ and\
  \bibinfo {author} {\bibfnamefont {R.}~\bibnamefont {Rattazzi}},\ }\bibfield
  {title} {\bibinfo {title} {{Gaugino mass without singlets}},\ }\href
  {https://doi.org/10.1088/1126-6708/1998/12/027} {\bibfield  {journal}
  {\bibinfo  {journal} {JHEP}\ }\textbf {\bibinfo {volume} {12}},\ \bibinfo
  {pages} {027}},\ \Eprint {https://arxiv.org/abs/hep-ph/9810442}
  {arXiv:hep-ph/9810442} \BibitemShut {NoStop}%
\bibitem [{\citenamefont {Pomarol}\ and\ \citenamefont
  {Rattazzi}(1999)}]{Pomarol:1999ie}%
  \BibitemOpen
  \bibfield  {author} {\bibinfo {author} {\bibfnamefont {A.}~\bibnamefont
  {Pomarol}}\ and\ \bibinfo {author} {\bibfnamefont {R.}~\bibnamefont
  {Rattazzi}},\ }\bibfield  {title} {\bibinfo {title} {{Sparticle masses from
  the superconformal anomaly}},\ }\href
  {https://doi.org/10.1088/1126-6708/1999/05/013} {\bibfield  {journal}
  {\bibinfo  {journal} {JHEP}\ }\textbf {\bibinfo {volume} {05}},\ \bibinfo
  {pages} {013}},\ \Eprint {https://arxiv.org/abs/hep-ph/9903448}
  {arXiv:hep-ph/9903448} \BibitemShut {NoStop}%
\bibitem [{\citenamefont {Davis}\ \emph {et~al.}(1983)\citenamefont {Davis},
  \citenamefont {Dine},\ and\ \citenamefont {Seiberg}}]{Davis:1983mz}%
  \BibitemOpen
  \bibfield  {author} {\bibinfo {author} {\bibfnamefont {A.~C.}\ \bibnamefont
  {Davis}}, \bibinfo {author} {\bibfnamefont {M.}~\bibnamefont {Dine}},\ and\
  \bibinfo {author} {\bibfnamefont {N.}~\bibnamefont {Seiberg}},\ }\bibfield
  {title} {\bibinfo {title} {{The Massless Limit of Supersymmetric {QCD}}},\
  }\href {https://doi.org/10.1016/0370-2693(83)91332-1} {\bibfield  {journal}
  {\bibinfo  {journal} {Phys. Lett. B}\ }\textbf {\bibinfo {volume} {125}},\
  \bibinfo {pages} {487} (\bibinfo {year} {1983})}\BibitemShut {NoStop}%
\bibitem [{\citenamefont {Dine}\ \emph {et~al.}(2017)\citenamefont {Dine},
  \citenamefont {Draper}, \citenamefont {Stephenson-Haskins},\ and\
  \citenamefont {Xu}}]{Dine:2016sgq}%
  \BibitemOpen
  \bibfield  {author} {\bibinfo {author} {\bibfnamefont {M.}~\bibnamefont
  {Dine}}, \bibinfo {author} {\bibfnamefont {P.}~\bibnamefont {Draper}},
  \bibinfo {author} {\bibfnamefont {L.}~\bibnamefont {Stephenson-Haskins}},\
  and\ \bibinfo {author} {\bibfnamefont {D.}~\bibnamefont {Xu}},\ }\bibfield
  {title} {\bibinfo {title} {{$\theta$ and the $\eta^\prime$ in Large $N$
  Supersymmetric QCD}},\ }\href {https://doi.org/10.1007/JHEP05(2017)122}
  {\bibfield  {journal} {\bibinfo  {journal} {JHEP}\ }\textbf {\bibinfo
  {volume} {05}},\ \bibinfo {pages} {122}},\ \Eprint
  {https://arxiv.org/abs/1612.05770} {arXiv:1612.05770 [hep-th]} \BibitemShut
  {NoStop}%
\bibitem [{\citenamefont {Cs\'aki}\ \emph
  {et~al.}(2023{\natexlab{b}})\citenamefont {Cs\'aki}, \citenamefont
  {Tito~D'Agnolo}, \citenamefont {Gupta}, \citenamefont {Kuflik}, \citenamefont
  {Roy},\ and\ \citenamefont {Ruhdorfer}}]{Csaki:2023yas}%
  \BibitemOpen
  \bibfield  {author} {\bibinfo {author} {\bibfnamefont {C.}~\bibnamefont
  {Cs\'aki}}, \bibinfo {author} {\bibfnamefont {R.}~\bibnamefont
  {Tito~D'Agnolo}}, \bibinfo {author} {\bibfnamefont {R.~S.}\ \bibnamefont
  {Gupta}}, \bibinfo {author} {\bibfnamefont {E.}~\bibnamefont {Kuflik}},
  \bibinfo {author} {\bibfnamefont {T.~S.}\ \bibnamefont {Roy}},\ and\ \bibinfo
  {author} {\bibfnamefont {M.}~\bibnamefont {Ruhdorfer}},\ }\bibfield  {title}
  {\bibinfo {title} {{On the dynamical origin of the $\eta'$ potential and the
  axion mass}},\ }\href {https://doi.org/10.1007/JHEP10(2023)139} {\bibfield
  {journal} {\bibinfo  {journal} {JHEP}\ }\textbf {\bibinfo {volume} {10}},\
  \bibinfo {pages} {139}},\ \Eprint {https://arxiv.org/abs/2307.04809}
  {arXiv:2307.04809 [hep-ph]} \BibitemShut {NoStop}%
\bibitem [{\citenamefont {Cs\'aki}\ \emph
  {et~al.}(2024{\natexlab{b}})\citenamefont {Cs\'aki}, \citenamefont
  {Ruhdorfer},\ and\ \citenamefont {Youn}}]{Csaki:2024lvk}%
  \BibitemOpen
  \bibfield  {author} {\bibinfo {author} {\bibfnamefont {C.}~\bibnamefont
  {Cs\'aki}}, \bibinfo {author} {\bibfnamefont {M.}~\bibnamefont {Ruhdorfer}},\
  and\ \bibinfo {author} {\bibfnamefont {T.}~\bibnamefont {Youn}},\ }\bibfield
  {title} {\bibinfo {title} {{Spontaneous CP breaking in a QCD-like theory}},\
  }\href {https://doi.org/10.1007/JHEP12(2024)066} {\bibfield  {journal}
  {\bibinfo  {journal} {JHEP}\ }\textbf {\bibinfo {volume} {12}},\ \bibinfo
  {pages} {066}},\ \Eprint {https://arxiv.org/abs/2407.06252} {arXiv:2407.06252
  [hep-ph]} \BibitemShut {NoStop}%
\bibitem [{\citenamefont {Jungnickel}\ and\ \citenamefont
  {Wetterich}(1998)}]{Jungnickel:1997yu}%
  \BibitemOpen
  \bibfield  {author} {\bibinfo {author} {\bibfnamefont {D.~U.}\ \bibnamefont
  {Jungnickel}}\ and\ \bibinfo {author} {\bibfnamefont {C.}~\bibnamefont
  {Wetterich}},\ }\bibfield  {title} {\bibinfo {title} {{The Linear meson model
  and chiral perturbation theory}},\ }\href
  {https://doi.org/10.1007/s100520050161} {\bibfield  {journal} {\bibinfo
  {journal} {Eur. Phys. J. C}\ }\textbf {\bibinfo {volume} {2}},\ \bibinfo
  {pages} {557} (\bibinfo {year} {1998})},\ \Eprint
  {https://arxiv.org/abs/hep-ph/9704345} {arXiv:hep-ph/9704345} \BibitemShut
  {NoStop}%
\bibitem [{\citenamefont {Ecker}\ \emph {et~al.}(1989)\citenamefont {Ecker},
  \citenamefont {Gasser}, \citenamefont {Pich},\ and\ \citenamefont
  {de~Rafael}}]{Ecker:1988te}%
  \BibitemOpen
  \bibfield  {author} {\bibinfo {author} {\bibfnamefont {G.}~\bibnamefont
  {Ecker}}, \bibinfo {author} {\bibfnamefont {J.}~\bibnamefont {Gasser}},
  \bibinfo {author} {\bibfnamefont {A.}~\bibnamefont {Pich}},\ and\ \bibinfo
  {author} {\bibfnamefont {E.}~\bibnamefont {de~Rafael}},\ }\bibfield  {title}
  {\bibinfo {title} {{The Role of Resonances in Chiral Perturbation Theory}},\
  }\href {https://doi.org/10.1016/0550-3213(89)90346-5} {\bibfield  {journal}
  {\bibinfo  {journal} {Nucl. Phys. B}\ }\textbf {\bibinfo {volume} {321}},\
  \bibinfo {pages} {311} (\bibinfo {year} {1989})}\BibitemShut {NoStop}%
\end{thebibliography}%

\end{document}